# Analytical solutions for Ising models on high dimensional lattices

Boris Kryzhanovsky, Leonid Litinskii, Vladislav Egorov

Scientific Research Institute for System Analysis Russian Academy of Sciences

kryzhanov@niisi.ras.ru, litin@mail.ru, rvladegorov@rambler.ru

**Abstract.** We use an *m*-vicinity method to examine Ising models on hypercube lattices of high dimensions $d \geq 3$. This method is applicable for both short-range and long-range interactions. We introduce a small parameter, which determines whether the method can be used when calculating the free energy. When we account for interaction with the nearest neighbors only, the value of this parameter depends on the dimension of the lattice $d$. We obtain an expression for the critical temperature in terms of the interaction constants that is in a good agreement with results of computer simulations. For $d = 5, 6, 7$, our theoretical estimates match the experiments both qualitatively and quantitatively. For $d = 3, 4$, our method is sufficiently accurate for calculation of the critical temperatures, however, it predicts a finite jump of the heat capacity at the critical point. In the case of the three-dimensional lattice ($d=3$), this contradicts to the commonly accepted ideas of the type of the singularity at the critical point. For the four-dimensional lattice ($d = 4$) the character of the singularity is under current discussion. For the dimensions $d = 1, 2$ the *m*-vicinity method is not applicable.

## I. INTRODUCTION

Statistical physics provides effective methods of analysis allowing us to investigate large systems of elementary "agents" and to determine macroscopic characteristics – in particular, the free energy based on interactions between the "agents". If we know the free energy – by means of computer simulations or theoretical calculations – we can calculate such properties of the system as the internal energy, magnetization, heat capacity, susceptibility. The singularities of the temperature dependences of these characteristics define critical temperatures at which the internal restructurings take place in the system and phase transitions occur.

Such results (even if they are not quite accurate) are important not only for physicists. In the second half of the 80s, the statistical physics methods were applied to estimate the storage capacity of the Hopfield neural network. Later on, a lot of investigations in the field of neural science based on the statistical physics have followed (see, for example, [1–3]). At the same time, the statistical physics methods became popular in the combinatorial optimization problems [4–6]. Starting from the mid-90s, a new scientific branch named econophysics appeared. In econophysics the statistical physics methods are the main instruments for analyzing economic models [7, 8].

Physics provides a wide variety of methods for the calculation of the free energy, from computer simulations (where one uses the Metropolis or the Wang-Landau algorithms [9-11]) to cumbersome theoretical approaches of the type of the renormalization-group or the transfer-matrix methods [12-14].

In the present paper, we sum up the results obtained when developing the *m*-vicinity method[1] for analysis of the Ising systems. Our method allows us to calculate the free energy for an arbitrary connection matrix. We present a review of our results for the Ising models on hypercubic lattice of the dimensions $d = 3, 4, 5, 6$, and $7$. The dimensions $d = 1$ and $d = 2$ are absent in this list because this method is not applicable for such lattices.

There is an enormous number of papers, which studied theoretically or experimentally the behavior of spin systems on specific types of lattices. Here we cite only the papers that we used in the course of our work. The cubic lattices are studied, for example, in [9, 15-19, 32, 33]. Four-dimension lattices were discussed in [20-22]. Studies on higher-dimensional systems are very rare. Our single source on $d \geq 5$ was [23].

In the next section, we justify the main approximation of our method. It consists of the substitution of the Gaussian distribution in place of the unknown state distribution of the given Hamiltonian.

In Sec. III, we show that the Gaussian approximation is the first order term in the expansion of the density of states in a perturbation theory series in a small parameter $\varepsilon_{max}$. In the case of the planar lattice ($d = 2$), $\varepsilon_{max} \approx 0.7$ and this value is not sufficiently small. When the lattice dimension $d$ increases, the value of $\varepsilon_{max}$ decreases quickly and the Gaussian approximation works very well.

---

[1] At first we called it "the *n*-vicinity method", but then it became clear that the more appropriate term was "the *m*-vicinity method".

In Sec. IV, we present in detail our results obtained with the aid of the Gaussian approximation of the density of states. The mean and the variance of the Gaussian distribution that we use coincide with the first and the second moments of the density of states of the given system. We define the boundaries of the method applicability and obtain analytical expressions for the critical characteristics of the system. In particular, they are the critical value of the inverse temperature and the jump of the heat capacity. Our analytical results match quite accurately with the results of computer simulations. We find out that the higher the dimension of the lattice the better our estimates; and when d > 4, the relative error is of the order of the tenth or the hundredth percent.

In Sec. V, we check whether the account of the second order terms of the perturbation theory improves our results for the critical temperature. Here we approximate the density of states with a distribution whose first three moments coincide with the moments of the state distribution of the system. We find that, while the role of second order parameters is negligible, we can achieve an almost perfect agreement with experiment by introducing an adjustable parameter.

In Sec. VI, we present a detail comparison of the theoretical results and computer simulations for the Ising model on the cubic lattice ($d = 3$). For such a lattice, we examine the role of the long-range interaction; in particular, we discuss the interactions with the next-nearest neighbors and the next-next-nearest neighbors.

Finally, in Sec. VII we sum up the strengths and weaknesses of the *m*-vicinity method. The details of the calculations are in Appendix.

## II. MAIN APPROXIMATION OF *m*-VICINITY METHOD

Let us examine the Ising model on a multidimensional cubic lattice, which is a system of $N$ spins $s_i = \{\pm 1\}$, $i = 1, 2, .., N$ situated at the nods of a hypercubic lattice. In what follows, we assume the periodic boundary conditions.

The Hamiltonian of the system is

$$E_H = E - mH, \quad E = -\frac{1}{2N}\sum_{i,j=1}^{N} J_{ij} s_i s_j, \quad m = \frac{1}{N}\sum_{i=1}^{N} s_i,$$

where $\mathbf{J} = (J_{ij})_1^N$ is a connection matrix, $H$ is a magnetic field, and $m$ is a magnetization of the state $\mathbf{s} = (s_1, s_2, ..., s_N)$. The partition function of the system is

$$Z = \sum_E D(E)\exp(-N\beta E_H),$$

where $\beta$ is the inverse temperature, the summation is carried out over all the values of the energy $E$ and $D(E)$ is the density of states (the degeneracy of the energy states).

In the general case, we do not know the energy distribution $D(E)$. It seems that we can define it with the aid of the central limit theorem. Indeed, the value of $E$ is the sum of $N(N-1)/2$ weakly connected random variables and the Gaussian distribution

$$2^N \sqrt{\frac{N}{2\pi\sigma_0^2}} \exp\left(-\frac{1}{2}N\frac{E^2}{\sigma_0^2}\right), \quad \text{where} \quad \sigma_0^2 = \frac{1}{2N}\sum_{i,j=1}^{N} J_{ij}^2, \tag{1}$$

describes correctly the central part of the distribution $D(E)$. However, Eq. (1) is not applicable at the tails of the true distribution, while the tails provide the main contribution to the formation of the phase transition. Many authors mentioned this fact (see [15]). The *m*-vicinity method allows us to overcome this difficulty. The essence of the method is as follows. We divide the set of $2^N$ configurations into $N+1$ subsets $\Omega_m$, which will be called the *m*-vicinities. The *m*-vicinity $\Omega_m$ contains all the configurations $\mathbf{s}$ with the same magnetization $m$. These configurations $\mathbf{s}$ differ from the configuration of the ground state $\mathbf{s}_0 = (1,1,...,1)$ by opposite signs of $n$ spins where $n = N(1-m)/2$ and the number of such configurations is equal to $\binom{N}{n}$.

The density $D(E,m)$ is the energy distribution for a given *m*-vicinity and the partition function of the system is

$$Z = \sum_{n=0}^{N}\sum_E D(E,m)\exp\left[-N\beta(E - mH)\right]. \tag{2}$$

In the general case, we do not know the true distribution $D(E,m)$. However, we do know the exact values of its mean energy $E_m$ and its variance, which we denote as $N^{-1}\sigma_m^2$ (see Appendix or [26-28]). In the case of the Ising model on the hypercube these expressions in the limit $N \to \infty$ are sufficiently simple

$$E_m = E_0 m^2, \quad N^{-1}\sigma_m^2 = N^{-1}\sigma_0^2 \cdot (1-m^2)^2, \quad E_0 = -\frac{1}{2N}\sum_{i,j=1}^{N} J_{ij}, \tag{3}$$

where $E_0 = E(\mathbf{s}_0)$ is the energy of the ground state $\mathbf{s}_0$ of the Ising Hamiltonian.

Let us explain the purpose of dividing the whole set of the configurations into $m$-vicinities. In the vicinity $\Omega_m$ the energy $E$ behaves as a random value and due to the central limit theorem we can approximate accurately the central part of $D(E,m)$ by Gaussian distribution with the mean $E_m$ and the variance $N^{-1}\sigma_m^2$:

$$D(E,m) = \frac{\sqrt{N}}{\sqrt{2\pi}\sigma_m} \binom{N}{n} \exp\left[-\frac{1}{2}N\left(\frac{E-E_m}{\sigma_m}\right)^2\right], \quad n = N(1-m)/2. \tag{4}$$

It is evident that the sum $\sum_m D(E,m)$ differs from the Gaussian distribution (1) and it better describes the tails of the true distribution.

Let us note here that in the limit $\sigma_m \to 0$ we can replace the exponent in Eq. (4) by a delta function $\delta(E - E_m)$. Then Eq. (2) takes the classical form known from the mean field theory, which provides the Bragg–Williams results [29]. However, the value of $\sigma_m$ differs from zero for all the types of the connection matrices (an exclusion is the case of the complete graph when all the matrix elements are equal – see Eq. (A4)) and the replacement of the Gaussian (4) by the delta function is not correct. It is the account for the value $\sigma_m \neq 0$ that leads to a much better agreement of theoretical estimates and results of simulations. Of course, the distribution $D(E,m)$ is not purely Gaussian, since its higher odd moments are not equal to zero (see Appendix), however, their contribution is sufficiently small. In what follows, we analyze when the Gaussian approximation (4) is applicable and how a deviation of the density of states $D(E,m)$ from Eq. (4) influences the results. In the next section, we show that the expression (4) is the first order term of the perturbation theory in a small parameter $(2q)^{-1/2}$, where $q$ is an effective number of neighbors (see Eq. (16)).

### III. SMALL PARAMETER IN $m$-VICINITY METHOD

The basis of the $m$-vicinity method is the abovementioned approximate description of the density of states. To analyze the approximation, we briefly repeat the calculations of the paper [30]. The starting point is as follows. We do not know a true energy distribution $D(E,m)$ but we do know the first moments of this distribution. In particular, we know the mean and the variance (3). Let us define the small parameter allowing us to expand the function $D(E,m)$ in a perturbation theory series.

We present $D(E,m)$ in the form

$$D(E,m) = \binom{N}{n}\exp\left[-N\varphi(m,E)\right],$$

where $\varphi = \varphi(m,E)$ is an unknown function and use the Stirling formula to replace the summation in (2) by integration. Then up to an insignificant constant, we obtain the partition function of the form

$$Z \sim \int_0^1 dm \int_{-\infty}^{\infty} dE \, e^{-NF(m,E)}, \tag{5}$$

where

$$F(m,E) = S(m) + \beta(E - mH) + \varphi(m,E),$$
$$S(m) = -\ln 2 + \frac{1}{2}\left[(1+m)\ln(1+m) + (1-m)\ln(1-m)\right]. \tag{6}$$

Let us estimate the integral (5) using the saddle point method. The equations for the saddle point are

$$\frac{\partial F}{\partial m} = \frac{1}{2}\ln\left(\frac{1+m}{1-m}\right) + \frac{\partial \varphi}{\partial m} - \beta H = 0, \qquad \frac{\partial F}{\partial E} = \frac{\partial \varphi}{\partial E} + \beta = 0. \qquad (7)$$

The solutions of these equations $m = M$ and $E = U$ are the spontaneous magnetization and the internal energy, respectively. Substituting these values in Eq. (6), we obtain the free energy $f(\beta) = F(M,U)$.

Now we turn to defining the small parameter of the *m*-vicinity method. Since the magnetic field $H$ does not influence the distribution $D(E,m)$ we set here $H = 0$. We write the function $\varphi(m,E)$ as a perturbation theory series in the vicinity of the point $E = E_m$:

$$\varphi = \frac{1}{2}\varepsilon^2 + \frac{1}{3!}\kappa_3\varepsilon^3 + \frac{1}{4!}\kappa_4\varepsilon^4 + \ldots , \quad \text{where} \quad \varepsilon = \frac{E - E_m}{\sigma_m} . \qquad (8)$$

The quantities $\kappa_k$ up to a sign coincide with the semi-invariants of the distribution $D(E,m)$ (see Appendix).

The main idea of the *m*-vicinity method is the possibility to restrict ourselves by accounting for just a few first terms of the series (8). We can do this only if $|\varepsilon| \ll 1$ and in this case the *m*-vicinity method is sufficiently accurate. Let us clarify that we are not interested in the values of $\varepsilon = \varepsilon(m,E)$ over the whole region of definition of the parameters $m$ and $E$. We have to know the expansion (8) only in a small vicinity of the saddle point, which is close to the values $m = M$ and $E = U$. Consequently, the small parameter we are looking for is

$$\varepsilon_0 = \frac{U - E_0 M^2}{\sigma_0(1 - M^2)} .$$

The smallness of the parameter $\varepsilon_0$ is the condition for applicability of the *m*-vicinity method.

In [30], there is a detailed analysis of the values of this parameter for different models. Here we restrict ourselves with an estimate of $\varepsilon_0$ by means of the reverse-reasoning method. Suppose the parameter $\varepsilon_0$ is small and it is sufficient to use only the first term of the series (8) and set $\varphi = \varepsilon^2/2$. Then the second of the equations (7) takes a simple form $\varepsilon = -\beta\sigma_m$ and we can rewrite the first of equations (7) as

$$\varepsilon\left(\varepsilon - \frac{E_0}{\sigma_0}\right) = -\frac{1-m^2}{4m}\ln\left(\frac{1+m}{1-m}\right) . \qquad (9)$$

We are interested in the behavior of the quantity $\varepsilon = \varepsilon_0$, where $\varepsilon_0$ is the root of the equation (9). The value of $|\varepsilon_0|$ reaches its maximum at the critical point $\beta = \beta_c$ (see [30]). In the limit $m \to 0$ ($\beta \to \beta_c$), Eq. (9) takes the form

$$\varepsilon^2 - \varepsilon\frac{E_0}{\sigma_0} + \frac{1}{2} = 0 .$$

The solution of this equation is

$$\varepsilon_{\max} = -\frac{|E_0|}{2\sigma_0} + \frac{1}{2}\sqrt{\frac{E_0^2}{\sigma_0^2} - 2} , \qquad (10)$$

and $\max|\varepsilon_0| = |\varepsilon_{\max}|$. Consequently, the *m*-vicinity method is applicable when the small parameter $|\varepsilon_0| \le |\varepsilon_{\max}|$. According to Eq. (10), it is necessary that $E_0^2/\sigma_0^2 \ge 2$ or $|\varepsilon_{\max}| \le 2^{-1/2}$.

A detailed analysis shows (see [27]) that a stricter inequality $E_0^2/\sigma_0^2 \ge 8/3$ or $|\varepsilon_{\max}| \le 6^{-1/2}$ defines the framework of the *m*-vicinity method. In other case, we obtain a jump of the spontaneous magnetization at the critical point and this contradicts to the known results.

We would like to mention here that when analyzing the dependence of our results on the lattice dimension $d$, we see that the ratio $E_0^2/\sigma_0^2$ depends on the effective number of the neighbors $q$ introduced below in Eq. (16). This parameter defines the number of interactions we take into account. We can rewrite Eq. (10) in terms of $q$; then

$\varepsilon_{\max} = \sqrt{q}\left(-1+\sqrt{1-4/q}\right)/2\sqrt{2}$. From this equation, it follows that when $q \gg 1$ the value of $|\varepsilon_{\max}| \sim 1/\sqrt{2q}$. Consequently, when $q$ increases, the value of the small parameter $|\varepsilon_{\max}|$ decreases quite rapidly and the accuracy of the $m$-vicinity method increases.

In Fig. 1, we show the dependence of $|\varepsilon_{\max}|$ on $d$ when we account for the nearest neighbors only (in this case $E_0^2/\sigma_0^2 = d$). For a planar lattice, $|\varepsilon_{\max}| \approx 0.7$. Obviously, it is problematic to use it as a small parameter. Moreover, for a planar lattice the inequality $E_0^2/\sigma_0^2 \geq 8/3$ is not satisfied. Next, for a cubic lattice the last inequality is satisfied and the value of $|\varepsilon_{\max}| \approx 0.36$ is more appropriate as a small parameter. Again, when $d$ increases the value of $|\varepsilon_{\max}|$ decreases and the accuracy of the method increases accordingly.

Summing up, we can say that the Gaussian approximation is the first order of the perturbation theory in the small parameter $|\varepsilon_{\max}|$.

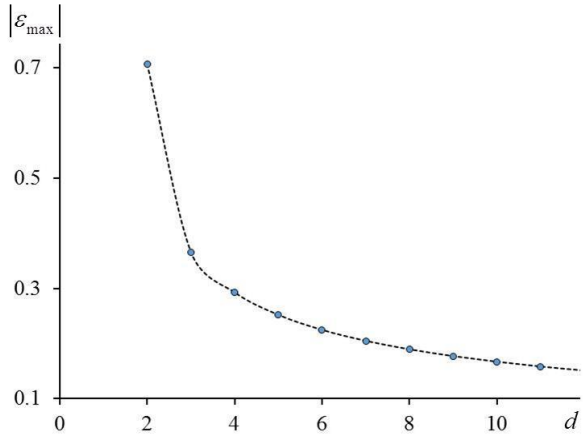

Fig. 1. Small parameter $|\varepsilon_{\max}|$ as function of dimension $d$ of Ising model with interaction of nearest neighbors.

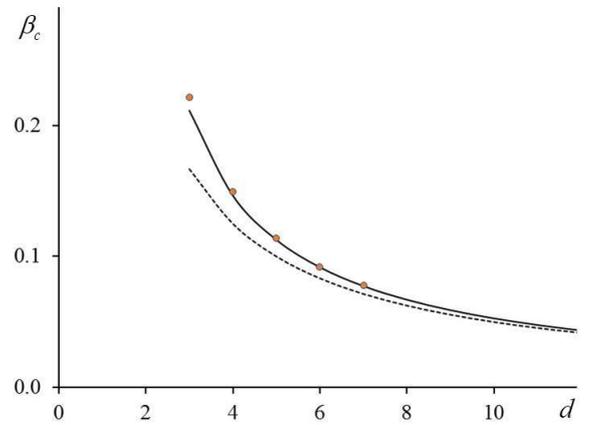

Fig. 2. Critical temperature $\beta_c$ vs $d$ and account for interactions with nearest neighbors only: solid line corresponds to Eq. (18); circles are simulations [9], [21], and [23]; dashed line is result of mean field theory.

## IV. FIRST ORDER OF PERTURBATION THEORY: GAUSSIAN APPROXIMATION

In this section, we use the Gaussian approximation (4) for the distribution $D(E,m)$. If in equations (6) – (8) we set $\varphi = \varepsilon^2/2$, the first equation (6) becomes

$$F(m,E) = S(m) + \beta(E - mH) + \frac{1}{2}\left(\frac{E-E_m}{\sigma_m}\right)^2; \qquad (11)$$

the equations for the saddle point (7) are

$$\frac{1}{2}\ln\left(\frac{1+m}{1-m}\right) = 2m\beta\left(|E_0| + \varepsilon\sigma_0\right) + \beta H, \qquad \varepsilon = -\beta\sigma_0(1-m^2). \qquad (12)$$

Eliminating the variable $\varepsilon$ from these equations, we obtain an equation of state

$$\frac{1}{2}\ln\left(\frac{1+m}{1-m}\right) = 2m\beta\left[|E_0| - \beta\sigma_0^2(1-m^2)\right] + \beta H. \qquad (13)$$

This equation differs from the well-known equation of Bragg and Williams [29] by a term proportional to $\beta^2$. After a transformation of Eq. (11) with the account for the equations (12) and (13) we obtain the expression for the free energy:

$$f(\beta) = \frac{1}{2}\ln\left(\frac{1-m^2}{4}\right) + \beta|E_0|m^2 - \frac{1}{2}\beta^2\sigma_0^2(1-m^2)(1+3m^2) \tag{14}$$

that defines $f$ as a function of $\beta$ and $m$. In Eq. (14) the spontaneous magnetization $m = m(\beta; H)$ is the solution of Eq. (13). Again by setting $\sigma_0^2 = 0$, we recover the known result from the mean field theory.

### A. Critical point

In this subsection, we set $H = 0$ and define the critical temperature under this assumption. For this purpose, we rewrite Eq. (13) as

$$\frac{1}{2m}\ln\left(\frac{1+m}{1-m}\right) = qb\left[1-b(1-m^2)\right] \tag{15}$$

where we introduce dimensionless characteristics

$$b = \beta\frac{\sigma_0^2}{|E_0|}, \quad q = \frac{2E_0^2}{\sigma_0^2} = \frac{\left(\sum_{i,j=1}^{N} J_{ij}\right)^2}{N\sum_{i,j=1}^{N} J_{ij}^2}. \tag{16}$$

The defined variables (16) are convenient since in Eq. (15) the only parameter that depends on the type of the lattice is $q$.

When $m \to 0$ ($\beta \to \beta_c$) the equation (15) takes the form

$$\frac{1}{q} = b(1-b) \tag{17}$$

Now, with account of Eq. (16) we obtain the expression for the critical temperature

$$\beta_c = b_c \frac{|E_0|}{\sigma_0^2}, \quad \text{where} \quad b_c = \frac{1}{2}\left(1 - \sqrt{1 - \frac{4}{q}}\right). \tag{18}$$

The critical value of $b_c$ depends only on the number $q$, and this parameter depends on the mean and the variance of the elements of the connection matrix. Since $q$ is a characteristic of the interactions that we take into account, we regard it as an effective number of the neighbors. In particular, if we account for an isotropic interaction with the nearest neighbors only, $\beta_c = b_c$, $q = 2d$, and $d$ is the dimension of the lattice. Note that in this case $q$ is exactly equal to the number of spins with which the given spin interacts. Then Eq. (18) describes pretty well the results of computer simulations for all the dimensions, which we examined (see Fig. 2 and Table 1.)

### B. Analytical expressions

Let us list the basic thermodynamic characteristics that we obtained from the equations (11) – (14).

**1)** The interval $\beta < \beta_c$. When $\beta < \beta_c$ and $m = 0$, from the above-mentioned equations we obtain that the free energy, the internal energy, and the heat capacity are

$$f = -\ln 2 - \frac{1}{2}\beta^2\sigma_0^2, \quad U = -\beta\sigma_0^2, \quad \text{and} \quad C = \beta^2\sigma_0^2, \tag{19}$$

respectively.

**2)** The interval $\beta \geq \beta_c$. To obtain the analytical expressions for the values $f$, $U$, and $C$ we solve the equation (15) for $b$:

$$b = \frac{1}{2(1-m^2)} \left[ 1 - \sqrt{1 - \frac{2(1-m^2)}{qm} \ln\left(\frac{1+m}{1-m}\right)} \right] \quad (20)$$

and transform the expressions (12) and (14) to the forms

$$f = S(m) - \frac{1}{2} qbm^2 - \frac{1}{4} qb^2 (1-m^2)^2, \quad (21)$$

$$U = E_0 \left[ m^2 + b(1-m^2)^2 \right], \quad (22)$$

$$\sigma_E^2 = \sigma_0^2 (1-m^2)^2 + 4E_0^2 \frac{m^2(1-m^2)\left[1 - 2b(1-m^2)\right]^2}{1 - qb(1-m^2)\left[1 - b(1-3m^2)\right]}. \quad (23)$$

Here $\sigma_E^2 = -d^2 f / d\beta^2$ is the energy variance that is related to the heat capacity via $C = \beta^2 \sigma_E^2$. The expression (23) is a result of differentiating the second equation (12) with respect to the variable $\beta$.

If we regard the magnetization as an independent variable $m \in [0,1]$ and consider the values of $f$, $U$, and $\sigma_E^2$ as functions of $m$, then Eqs. (20 – 23) define implicitly the dependence of $f$, $U$, and $\sigma_E^2$ on the inverse temperature $\beta = b|E_0|/\sigma_0^2$ at the interval $\beta \geq \beta_c$.

### C. Critical parameters

To examine the behavior of the thermodynamic characteristics near the critical point we introduce a relative inverse temperature

$$t = \frac{\beta - \beta_c}{\beta_c}.$$

Omitting intermediate calculations, we only present the most important critical dependences.

**1)** From the equations (19 -23) it follows that at $\beta = \beta_c$ the free energy and the internal energy are continuous functions and the heat capacity has a jump. Indeed, when $\beta > \beta_c$ the equation (23) holds, and in the limit $m \to 0$ ($\beta \to \beta_c$) we obtain

$$\sigma_E^2 = \sigma_0^2 + 6E_0^2 \frac{(1-2b_c)^2}{1 - 3qb_c^2}, \quad (24)$$

where $b_c$ is defined by Eq. (18). Comparing this expression with Eq. (19) we see that at $\beta = \beta_c$ the energy variance has a jump and consequently the heat capacity also has a jump:

$$\Delta C = \frac{3}{2} q^2 b_c^2 \frac{(1-2b_c)^2}{1 - 3qb_c^2}. \quad (25)$$

For the lattice dimensions $d = 5$, 6, and 7 the authors of [23] used computer simulations to estimate the jumps of the heat capacity. The comparison of their results with the values following from Eq. (25) shows:

$$d = 5, \quad \Delta C^{(\exp)} = 1.8703, \quad \Delta C^{(theor)} = 1.8469$$
$$d = 6, \quad \Delta C^{(\exp)} = 1.7403, \quad \Delta C^{(theor)} = 1.7394$$
$$d = 7, \quad \Delta C^{(\exp)} = 1.6860, \quad \Delta C^{(theor)} = 1.6824.$$

We see that the formula (25) provides a very good agreement with the computer simulations; the larger $d$ the better this agreement.

**2)** When $\beta > \beta_c$, the value of the spontaneous magnetization near the critical point ($t \to 0$) obtained from Eq. (15) is

$$M = A\sqrt{t}, \quad A = \left( \frac{1 - qb_c^2}{\frac{1}{3} - qb_c^2} \right)^{\frac{1}{2}}. \quad (26)$$

This expression differs from the dependence $M \sim t^{\frac{1}{8}}$ that is valid for the two-dimensional Ising model ($q=4$), however, it qualitatively coincides with the expressions $M = 2\sqrt{t}$ and $M = \sqrt{3t}$ obtained in the framework of the van der Waals theory [13] and the mean field theory, respectively. Note that our result tends to the result of the mean field theory in the limit $q \gg 1$. It is also worthwhile to note that Eq. (26) predicts a larger magnetization than the mean field theory, $A > \sqrt{3}$ for any $q \geq 4$. If $q \leq 11$ ($A > 2$) the van der Waals magnetization is less than the value (26) and when $q > 11$ ($A < 2$) it is larger than the value (26).

**3)** From Eq. (13), it follows that at the critical point the susceptibility has a jump

$$\chi^{-1} = \begin{cases} -qt\sqrt{1-4/q} & , \quad t < 0 \\ 2qt\sqrt{1-4/q} & , \quad t > 0. \end{cases} \tag{27}$$

Comparing with the analogous expression of the mean field theory, we see that an extra factor $\sqrt{1-4/q}$ appears in Eq. (27), and it tends to 1 when $q \gg 1$. As opposed to the mean field theory [13], in Eq. (27) $q$ is the effective number of the neighbors (16).

**4)** It is easy to see that our model satisfies the similarity hypothesis. Indeed, when we expand the expression (13) in small parameters $m$ и $t$, we obtain the dependence $H = H(m,t)$ that can be rewritten in the classical form $\beta_c H = m|m|^{\delta-1} h_s(tm^{-2})$ with the critical exponent $\delta = 3$ and the scaling function

$$h_s(x) = \left(\frac{1}{3} - qb_c^2\right) - x\left(1 - qb_c^2\right) . \tag{28}$$

### D. Magnetization distribution

The integral

$$P(m) = Z^{-1} \int_{E_0}^{|E_0|} D(E,m) dE \tag{29}$$

defines the probability of finding the system in a state with the magnetization $m$. In the Gaussian approximation we use here, $D(E,m)$ is defined by Eq. (4). We estimate this integral with the aid of the saddle point method. The value of $P(m)$ is accurate to a normalization constant $P_0$

$$P(m) = P_0 e^{-N\Phi(m)}, \quad \text{where} \quad \Phi(m) = S(m) + \beta E_m - \frac{1}{2}\beta^2 \sigma_m^2. \tag{30}$$

In Fig. 6a, we show the typical behavior of the curves (30). As we might expect, after crossing the critical point the bimodal distribution replaces the unimodal distribution.

We can use Eq. (30) when analyzing the Binder cumulant $Q = 1 - \langle m^4 \rangle / 3\langle m^2 \rangle^2$ (see [31]). In Fig. 6b, we show the curves $Q = Q(\beta)$ for cubic lattices whose linear sizes are $L = 8, 10,$ and 12. We are interested in the value of the cumulant $Q_c = Q(\beta_c)$ at the critical point. To calculate $Q_c$ we use the Taylor expansion of the function $\Phi(m)$ in Eq. (30):

$$\Phi(m) \approx -\frac{1}{2}\beta^2 \sigma_0^2 + \frac{1}{2!} a_2 m^2 + \frac{1}{2!} a_4 m^4 , \tag{31}$$

where

$$a_2 = \left.\frac{d^2\Phi}{dm^2}\right|_{m=0} = 1 + 2\beta E_0 + 2\beta^2 \sigma_0^2, \qquad a_4 = \left.\frac{d^4\Phi}{dm^4}\right|_{m=0} = 1 - 12\beta^2 \sigma_0^2 .$$

This expansion is useful when $\beta \leq \beta_c + O(N^{-1/2})$ and the value of $P(m)$ is noticeably nonzero only when $m \ll 1$. In this case, the distribution (30) takes the form

$$P(m) = P_0 \exp\left[-\frac{1}{2!}Na_2 m^2 - \frac{1}{4!}Na_4 m^4\right],$$

which allows us to calculate the normalization constant $P_0$ easily as well as the mean values $\langle m^2 \rangle$ and $\langle m^4 \rangle$. It is not difficult to see that when $|\beta - \beta_c| > O(N^{-1/2})$ we can consider the distribution $P(m)$ as purely Gaussian in full agreement with the results of [31]. However, when $\beta \to \beta_c$ we have $a_2 \to 0$ and at the critical point the distribution takes the form $P(m) = P_0 \exp\left[-Na_4 m^4 / 4!\right]$. Using this expression to calculate the critical value of the Binder cumulant [31] in the limit $N \to \infty$ we obtain

$$Q_c = 1 - \frac{\Gamma(5/4) \cdot \Gamma(1/4)}{3\Gamma(3/4)^2} \approx 0.2705. \tag{32}$$

It is interesting that in the limit $N \to \infty$ the value $Q_c$ does not depend on any parameters of the model (the lattice type, the character of the long-range interaction, and so on). Although we obtained this result in the framework of the Gaussian approximation (4), it has a general character. Indeed, let $D(E,m)$ be an unknown function and the integration (29) results in the expression (30) where $\Phi(m)$ is also an unknown function. By a symmetry argument, it follows that only even powers of $m$ are present in the Taylor expansion of this function. At the critical point the second derivative of the function $\Phi(m)$ equals to zero (the bimodal distribution replaces the unimodal) and the Taylor series starts from the term $\sim m^4$. We assume that the function $\Phi(m)$ has no singularities and its derivatives are finite. Then we can leave only the term $\sim m^4$ in the exponent of the distribution $P(m)$. The reason is that when integrating over $m$ the account for the terms of the higher orders leads to corrections of the order $N^{-1/2}$. In other words, we can present the magnetization distribution as $P(m) = P_0 \exp\left[-Na_4 m^4 / 4!\right]$ where $a_4$ is an unknown, which is canceled out in the calculation of $Q_c$ and do not influence the final form of the expression (32).

### E. Density of states

In Subsec. IV.2 we derived the equations (19) – (23), which allowed us to obtain implicitly the logarithmic density of states

$$\Psi(E) = \beta E - f(\beta); \quad E = df/d\beta$$

using the Legendre relations. When $\beta$ changes from 0 to $\infty$, the value of $E$ changes from 0 to $E_0$ and for each $\beta$ we obtain a pair of values $E$ and $\Psi(E)$. In such a way we generate the function $\Psi(E)$, which we suppose to be symmetric: $\Psi(-E) = \Psi(E)$. In Fig. 7a, we present the comparison of our results with computer simulations.

Let us determine an explicit form of the dependence $\Psi = \Psi(E)$. The integral

$$D(E) = \int_{-1}^{1} D(E,m) dm \approx c_1 \cdot \int_{-1}^{1} e^{-N\Lambda(m,E)} dm = c_2 e^{N\Psi(E)}$$

defines the density of states or, in other words, the number of states with the energy $E$. Here $\Lambda(m,E) = S(m) + (E - E_m)^2 / 2\sigma_m^2$; $c_1$ and $c_2$ are non-essential constants. Therefore, the logarithmic density of states is $\Psi(E) = \Lambda(m_E, E)$, where $m_E = m_E(E)$ is the saddle point, which is a solution of the equation $\partial \Lambda(m,E)/\partial m = 0$ After some transformations we can write this equation as

$$\frac{(1-m_E^2)^3}{2m_E} \ln\left(\frac{1+m_E}{1-m_E}\right) = q\left(1 - \frac{E}{E_0}\right)\left(\frac{E}{E_0} - m_E^2\right). \tag{33}$$

From this equation, it follows that

$$\Psi(E) = \begin{cases} \ln 2 - \dfrac{E^2}{2\sigma_0^2}, & \text{when } E \geq E_0 b_c \\ -S(m_E) - \dfrac{1}{2\sigma_0^2}\left(\dfrac{E - E_0 m_E^2}{1 - m_E^2}\right)^2, & \text{when } E < E_0 b_c, \end{cases} \tag{34}$$

where $b_c$ is defined by Eq. (18) and

$$m_E^2 \approx 1 - 2r \cdot \cos\varphi_E, \quad r = \sqrt{\frac{q}{3}\left(1 - \frac{E}{E_0}\right)}, \quad \text{and} \quad \varphi_E = \frac{1}{3}\left[\pi + \arccos\left(\frac{9r}{2q}\right)\right]. \tag{35}$$

The equation (35) is an approximate result obtained taking into account that the left-hand side of Eq. (33) contributes significantly to the solution of this equation only when $m_E \ll 1$. Under this condition $\ln\left[(1+m_E)/(1-m_E)\right]/2m_E \approx 1$ and Eq. (33) reduces to a cubic equation for the quantity $(1-m_E^2)$. The solution of this equation has the form (35). The approximate solution (34) differs by fractions of a percent from the exact solution obtained by means of the Legendre relations (see Fig. 7b).

Table 1. Critical temperatures and relative errors for the lattices dimensions $3 \le d \le 7$. Exact values are computer simulations [9], [21], and [23] rounded to 5 decimal places. For comparison, results of mean field theory also shown.

|  | Critical temperature $\beta_c$ | | | | |
|---|---|---|---|---|---|
|  | $d=3$ | $d=4$ | $d=5$ | $d=6$ | $d=7$ |
| Exact value | 0.22166 | 0.14970 | 0.11392 | 0.09230 | 0.07771 |
| Account for third moment, Eq. (40), $k=-0.8$ | 0.22155 | 0.14894 | 0.11372 | 0.09227 | 0.07772 |
| Gaussian approximation (18), Eq. (40), $k=0$ | 0.21132 | 0.14645 | 0.11270 | 0.09175 | 0.07742 |
| Account for third moment, Eq. (40), $k=1$ | 0.20196 | 0.14366 | 0.11151 | 0.09113 | 0.07706 |
| Mean field theory | 0.16667 | 0.12500 | 0.10000 | 0.08333 | 0.07143 |
|  | Relative error $Err$ | | | | |
| Account for third moment, Eq. (40), $k=-0.8$ | 0.02% | 0.25% | 0.09% | 0.02% | 0.01% |
| Gaussian approximation (18), Eq. (40), $k=0$ | 2.39% | 1.10% | 0.54% | 0.30% | 0.18% |
| Account for third moment, Eq. (40), $k=1$ | 4.65% | 2.06% | 1.07% | 0.64% | 0.42% |
| Mean field theory | 14.16% | 8.99% | 6.50% | 5.10% | 4.21% |

## V. SECOND ORDER OF PERTURBATION THEORY: ACCOUNT FOR THIRD MOMENT

In this section, we analyze what happens if, when approximating the distribution $D(m,E)$, we account for the third moment. We will examine only the simplest case supposing that all the elements of the connection matrix are equal ($J_{ij} = J$). Otherwise, the obtained expressions are too cumbersome and it is very difficult to analyze them. We restrict ourselves only by first two terms of the series expansion (8) and set

$$\varphi = \frac{1}{2}\varepsilon^2 + \frac{1}{3!}\kappa_3 \varepsilon^3, \quad \text{where} \quad \varepsilon = \frac{E - E_m}{\sigma_m}. \tag{36}$$

We define the coefficient $\kappa_3$ from the following considerations. It is necessary that the third moment of our approximation $\exp[-N \cdot \varphi(m,E)]$ coincides with the third moment of the true distribution $D(m,E)$: $\int (E-E_m)^3 \exp[-N \cdot \varphi(m,E)] dE = \mu_3(m)$. In Appendix, we show that for this it is necessary that the coefficient $\kappa_3$ is equal to the third semi-invariant of the distribution $D(m,E)$: $\kappa_3 = -\mu_3(m)/\sigma_m^3$. In Appendix we also obtain the expression for the third moment of the distribution $D(m,E)$, which is equal to $\mu_3(m) = -2qm^2\left(1-m^2\right)^2$. Here $q$ is the effective number of the neighbors (see Eq. (16)).

Then

$$\kappa_3 = \frac{2qm^2}{\sigma_0^3(1-m^2)},$$

and the expression (6) for the function $F(m, E)$ takes the form ($H = 0$):

$$F(m, E) = S(m) + \beta E + \frac{1}{2}\varepsilon^2 + \frac{1}{3!}\kappa_3 \varepsilon^3.$$

We recall that $E_m = E_0 m^2$, $\sigma_m = \sigma_0(1-m^2)$, $E_0 = -q/2$, and $\sigma_0 = \sqrt{q/2}$.

The system of equations that define the saddle point has the form

$$\frac{\partial F}{\partial E} = \beta + \frac{\varepsilon + \kappa_3 \varepsilon^2/2}{\sigma_m} = 0,$$
$$\frac{\partial F}{\partial m} = \frac{1}{2}\ln\frac{1+m}{1-m} + \left(\varepsilon + \frac{1}{2}\kappa_3 \varepsilon^2\right)\dot\varepsilon + \frac{\dot\kappa_3}{3!}\varepsilon^3 = 0, \quad (37)$$

where $\dot\varepsilon = \partial \varepsilon/\partial m$. The first of the equations (37) provides the relation

$$\varepsilon + \kappa_3 \frac{\varepsilon^2}{2} = -\beta\sigma_m. \quad (38)$$

Moreover, by direct calculations we obtain

$$\dot\varepsilon = \frac{2m}{1-m^2}\left(\varepsilon + \frac{|E_0|}{\sigma_0}\right), \quad \dot\kappa_3 = \frac{4qm}{\sigma_0 \sigma_m^2},$$

and substituting these equalities in the second equation (37), we finally have

$$\frac{1}{2m}\ln\frac{1+m}{1-m} = 2\beta\left(\varepsilon\sigma_0 + |E_0|\right) - \frac{2q}{3\sigma_0 \sigma_m^2}\varepsilon^3. \quad (39)$$

Generally speaking, for deriving an equation relating $m$ and $\beta$ it is necessary to solve Eq. (38) and to determine $\varepsilon = \varepsilon(m)$ and substitute it into Eq. (39). Analyzing this equation, it would be possible to define the region of applicability of the *m*-vicinity approximation with account for the third moment and to obtain an expression for the critical temperature. It is rather difficult to solve this problem analytically. This is the reason why we restrict ourselves by analyzing the influence of the third moment $\mu_3(m)$ on the value of the critical temperature defined by Eq. (39) when $m = 0$.

When $m \to 0$, from Eq. (38) it follows that $\varepsilon \to -\beta\sigma_0$. Then the second term in the right-hand side of Eq. (39) tends to $-2q\beta^3/3$ and this equation itself takes the form $1 = 2\beta(-\beta\sigma_0^2 + |E_0|) + 2q\beta^3/3$. By virtue of the expressions for $E_0$ and $\sigma_0$ we obtain the equation for the critical value of the inverse temperature $\beta_c$:

$$\frac{1}{q} = \beta - \beta^2 + \frac{2k}{3}\beta^3, \text{ where } k = 1. \quad (40)$$

If in this equation we set $k = 0$, we obtain Eq. (17) that corresponds to the above-discussed case of the Gaussian approximation. We introduce the parameter $k$ since in what follows we will use it as an adjustable parameter.

Solving Eq. (40) for $d = 3 \div 7$ numerically we see the worse agreement of the obtained results with the experiments (see Table 1). Previously (see [24, Subsection 17.6]) it was mentioned that not always an account for higher moments led to an increase of the quality of an approximation. You can expect such a result only for a narrow class of distributions. This question was under analysis when studying the Gram-Charlier and the Edgeworth series expansions [24, 25]. In particular, it turned out that when we omit the higher order terms of the infinite series (8), the error is of the order of the first omitted term. To compensate this error, we use the adjustable parameter $k$. The author of [24, s.17.6] suggests to use this receipt if the optimal value of $k$ gives a better agreement with all the experimental values. The questions of convergence of an infinite series are of secondary importance when solving a specific problem.

With those arguments in mind, we found that an agreement with the experimental results is better when $k < 0$. In this case, the solution of Eq. (40) takes the form

$$\beta_c = B\cos\phi - \frac{1}{2|k|}, \qquad (41)$$

where

$$B = \frac{\sqrt{1+2|k|}}{|k|}, \quad \phi = \frac{1}{3}(2\pi - \arccos R), \quad R = -\frac{1}{(1+2|k|)^{3/2}}\left(1+3|k|+\frac{6k^2}{q}\right).$$

For $d = 3$ the best agreement is reached when $k = -0.8$. It turned out that this value of $k$ is universal: the solutions of Eq. (41) with the same adjustable parameter provide a very good agreement with the experimental data for the lattices of all the dimensions $3 \leq d \leq 7$ ($q = 2d$). As we see from Table 1, account for the third moment and introduction of the adjustable parameter $k = -0.8$ reduce the relative error

$$Err = 100\% \cdot \left(\frac{\beta_c^{(\exp)} - \beta_c^{(theor)}}{\beta_c^{(\exp)} + \beta_c^{(theor)}}\right)$$

by 1 - 2 orders of magnitude, and it becomes comparable with the experimental error. When we account for the interactions with the second and third neighbors, the introduction of the third moment also improves the agreement with the experiment significantly (see the next section).

## VI. COMPARISON WITH EXPERIMENT: THREE-DIMENSIONAL ISING MODEL

To estimate the accuracy and the correctness of the obtained equations we performed computer simulations using the Metropolis algorithm and the algorithm of Wang and Landau [11]. We restricted ourselves to the examination of the three-dimensional Ising model supposing that for lattices of higher dimensions ($d \geq 4$) the agreement with the experiment would be only better. In the course of our experiment, we explored the functions $f = f(\beta)$, $U = U(\beta)$, and $C = C(\beta)$. This allowed us to define the dependence of the critical temperature $\beta_c = \beta_c(q)$ on $q$, to calculate the logarithmic density of states $\Psi(E)$, and to analyze how the magnetization distribution changed when the inverse temperature increased.

Since recently, there is a strong interest in an account for interactions with the second and the third neighbors (see, for example, [32] and [33]). However, we do not know any analytical estimates for the critical temperatures. Our Eq. (18) is quite accurate for such problems too.

We use the Metropolis algorithm to examine the three-dimensional lattices of the linear sizes $L = 20$, 32, and 64 with periodic boundary conditions. We look for the dependences of the critical parameters on the effective number of neighbors $q$. In our calculations, we suppose that all the interaction constants with the 6 nearest neighbors are equal to one ($J_{NN} = 1$). We start with varying the value of the constant of interaction with the 12 next-nearest neighbors $J_{NNN}$ from 0 to 1, and this means that $q$ changes from 6 to 18. Then we fix the value $J_{NNN} = 1$ and vary the interaction constant with the 8 next-next-nearest neighbors $J_{NNNN}$ from 0 to 1 so that $q$ changes from 18 to 26. The initial state of the system was random. To bring the system to a state close to equilibrium for a given temperature, at the first $10^4 \cdot L^3$ Monte-Carlo steps we do not accumulate the statistical data. We flip spins according the Metropolis algorithm in the order of their sequence in the lattice. After each flip, we measure the values of the energy and the magnetization. The total number of the Monte-Carlo steps is $4 \cdot 10^5 \cdot L^3$. We vary the inverse temperature with the step size $\Delta\beta \approx 1.8 \cdot 10^{-4}$ for L=20, with the step size $\Delta\beta \approx 7.5 \cdot 10^{-6}$ for L=32, and with the step size $\Delta\beta \approx 10^{-4}$ for L=64. We define the critical point as the point of maximum of the energy variance $\sigma_E^2 = \sigma_E^2(\beta)$. At this point we fix the values of $U_{\max} = U(\beta_{\max})$, $\sigma_{\max}^2 = \sigma_E^2(\beta_{\max})$ as well as other values. We also use the Metropolis algorithm to calculate the energy and magnetization moments near the critical temperatures. In Table 2, we present the experimental data for $L = 64$.

This calculation had two goals. The first was to analyze the role of the next terms of the expansion series (8). The second goal was to examine the influence of the account for the next-nearest and the next-next-nearest neighbors on the character of the dependence $C = C(\beta)$ and find if the type of the singularity at the critical point changed to a finite jump.

### A. Dependence $\beta_c = \beta_c(q)$

In Fig. 3a, we show the dependence of the $\beta_c$ on $q$. We compare the experimental values (the upper solid line) with those that follow from Eq. (18) for the Gaussian approximation (the lower solid line). As we see, the solid curves have similar shapes, and when we scale the theoretical curve by a factor $\sim 1.06$, they practically merge. Such a coincidence cannot be accidental. From Eq. (18) it follows that in the Gaussian approximation the curve $\beta_c(q)$ have singularities at $q = 18$ and $q = 26$:

$$\left.\frac{d\beta_c}{dq}\right|_{q\to 18-0} = -\frac{\beta_c}{6\sqrt{18-q}} \quad \text{and} \quad \left.\frac{d\beta_c}{dq}\right|_{q\to 26-0} = -\frac{\beta_c}{3\sqrt{26(26-q)}}.$$

The experimental curve shows the same singularities. In our simulations, we check this statement very accurately changing $q$ in the vicinities of these points with a very small step size ($\Delta q \sim 10^{-3}$).

Table 2. Values measured at $\beta = \beta_{\max}$.

| $J_{NNN}$ | $J_{NNNN}$ | $q$ | $\beta_{\max}$ | $\sigma^2_{\max}$ | $|\langle E \rangle|$ | $\langle E^2 \rangle$ | $\langle E^4 \rangle$ | $|\langle m \rangle|$ | $\langle m^2 \rangle$ | $\langle m^4 \rangle$ | $Q_{\max}$ |
|---|---|---|---|---|---|---|---|---|---|---|---|
| 0.0000 | 0 | 6.00 | 0.22208 | 71.02 | 1.0247 | 1.0502 | 1.1041 | 0.06448 | 0.04742 | 0.00258 | 0.6173 |
| 0.0042 | 0 | 6.10 | 0.21978 | 74.45 | 1.0270 | 1.0550 | 1.1143 | 0.10751 | 0.04862 | 0.00271 | 0.6173 |
| 0.0410 | 0 | 7.00 | 0.20058 | 92.80 | 1.0423 | 1.0866 | 1.1823 | 0.11810 | 0.05241 | 0.00310 | 0.6243 |
| 0.0811 | 0 | 8.00 | 0.18328 | 107.06 | 1.0517 | 1.1065 | 1.2260 | 0.14503 | 0.04908 | 0.00273 | 0.6223 |
| 0.1213 | 0 | 9.00 | 0.16888 | 132.16 | 1.0623 | 1.1290 | 1.2770 | 0.08342 | 0.04356 | 0.00223 | 0.6084 |
| 0.1623 | 0 | 10.00 | 0.15648 | 129.91 | 1.0834 | 1.1743 | 1.3813 | 0.17143 | 0.04511 | 0.00229 | 0.6252 |
| 0.2048 | 0 | 11.00 | 0.14548 | 154.83 | 1.1020 | 1.2151 | 1.4792 | 0.15042 | 0.04237 | 0.00206 | 0.6184 |
| 0.2500 | 0 | 12.00 | 0.13548 | 189.87 | 1.1300 | 1.2776 | 1.6360 | 0.10335 | 0.04287 | 0.00212 | 0.6154 |
| 0.2991 | 0 | 13.00 | 0.12608 | 199.76 | 1.1546 | 1.3339 | 1.7833 | 0.01969 | 0.03976 | 0.00183 | 0.6145 |
| 0.3543 | 0 | 14.00 | 0.11709 | 243.49 | 1.1972 | 1.4342 | 2.0623 | 0.07870 | 0.04215 | 0.00203 | 0.6194 |
| 0.4189 | 0 | 15.00 | 0.10799 | 287.88 | 1.2179 | 1.4844 | 2.2101 | 0.06442 | 0.03106 | 0.00123 | 0.5765 |
| 0.5000 | 0 | 16.00 | 0.09869 | 325.40 | 1.3223 | 1.7497 | 3.0701 | 0.08284 | 0.04788 | 0.00249 | 0.6373 |
| 0.6169 | 0 | 17.00 | 0.08769 | 426.66 | 1.4011 | 1.9647 | 3.8730 | 0.03163 | 0.03256 | 0.00142 | 0.5536 |
| 0.7106 | 0 | 17.50 | 0.08059 | 482.31 | 1.4871 | 2.2134 | 4.9152 | 0.12453 | 0.04236 | 0.00200 | 0.6290 |
| 0.8566 | 0 | 17.90 | 0.07149 | 586.79 | 1.5548 | 2.4198 | 5.8771 | 0.02487 | 0.02444 | 0.00080 | 0.5519 |
| 0.8958 | 0 | 17.95 | 0.06949 | 600.78 | 1.6359 | 2.6783 | 7.1979 | 0.03046 | 0.03645 | 0.00151 | 0.6216 |
| 0.9324 | 0 | 17.98 | 0.06769 | 655.33 | 1.6793 | 2.8225 | 7.9946 | 0.03795 | 0.03277 | 0.00129 | 0.5992 |
| 1.0000 | 0.0000 | 18.00 | 0.06448 | 736.77 | 1.6900 | 2.8587 | 8.2047 | 0.01656 | 0.02573 | 0.00087 | 0.5643 |
| 1 | 0.0063 | 18.10 | 0.06429 | 768.80 | 1.6916 | 2.8644 | 8.2385 | 0.06596 | 0.03199 | 0.00123 | 0.5993 |
| 1 | 0.0637 | 19.00 | 0.06248 | 770.88 | 1.7294 | 2.9938 | 8.9982 | 0.07366 | 0.03418 | 0.00136 | 0.6133 |
| 1 | 0.1307 | 20.00 | 0.06049 | 848.33 | 1.7475 | 3.0569 | 9.3844 | 0.03053 | 0.03480 | 0.00140 | 0.6135 |
| 1 | 0.2023 | 21.00 | 0.05848 | 904.70 | 1.7442 | 3.0458 | 9.3187 | 0.08452 | 0.03413 | 0.00135 | 0.6132 |
| 1 | 0.2806 | 22.00 | 0.05649 | 926.04 | 1.7881 | 3.2008 | 10.2901 | 0.18149 | 0.03500 | 0.00150 | 0.5918 |
| 1 | 0.3693 | 23.00 | 0.05439 | 989.95 | 1.8190 | 3.3125 | 11.0228 | 0.08570 | 0.04307 | 0.00201 | 0.6390 |
| 1 | 0.4755 | 24.00 | 0.05199 | 1060.91 | 1.7446 | 3.0478 | 9.3385 | 0.00448 | 0.02501 | 0.00080 | 0.5762 |
| 1 | 0.6177 | 25.00 | 0.04919 | 1157.27 | 1.7578 | 3.0944 | 9.6306 | 0.01029 | 0.02170 | 0.00063 | 0.5537 |
| 1 | 0.7225 | 25.50 | 0.04739 | 1324.75 | 1.8598 | 3.4641 | 12.0698 | 0.03932 | 0.03389 | 0.00131 | 0.6203 |
| 1 | 0.8707 | 25.90 | 0.04499 | 1418.32 | 1.8758 | 3.5239 | 12.4942 | 0.09043 | 0.02928 | 0.00101 | 0.6080 |
| 1 | 0.9076 | 25.95 | 0.04449 | 1414.33 | 1.9694 | 3.8839 | 15.1680 | 0.20027 | 0.04111 | 0.00182 | 0.6402 |
| 1 | 0.9410 | 25.98 | 0.04399 | 1455.97 | 1.9562 | 3.8324 | 14.7722 | 0.15120 | 0.03737 | 0.00154 | 0.6334 |
| 1 | 1.0000 | 26.00 | 0.04309 | 1505.02 | 1.8808 | 3.5433 | 12.6366 | 0.01037 | 0.02359 | 0.00071 | 0.5765 |

Agreement of the experiment with our theory becomes much better when we account for the third moment in Eq. (36) and solve numerically the equation for the saddle point (37) (accounting for Eqs. (A4) and (A11) of Appendix). In this case, the relative error does not exceed a fraction of a percent. When we account for interactions beyond the nearest neighbors, the expression for the third moment $\mu_3(m)$ becomes so cumbersome that does not allow us to obtain analytical expressions for the critical temperature analogous to Eqs. (40) and (41). On the other hand, we can use a simple empirical formula

$$\beta_c = \frac{\beta_c^{(0)}}{1+0.207\cdot(12 J_{NNN} + 8 J_{NNNN})}, \qquad (42)$$

where $\beta_c^{(0)} \approx 0.22165$ is the critical value given by formula (41) with account for the nearest neighbors only ( $J_{NNN} = J_{NNNN} = 0$ ). Comparing the solid lines in Fig. 3b, we see that the expression (42) describes the experiment much better than formula (18) - in comparison with the Gaussian approximation the relative error decreases by an order of magnitude.

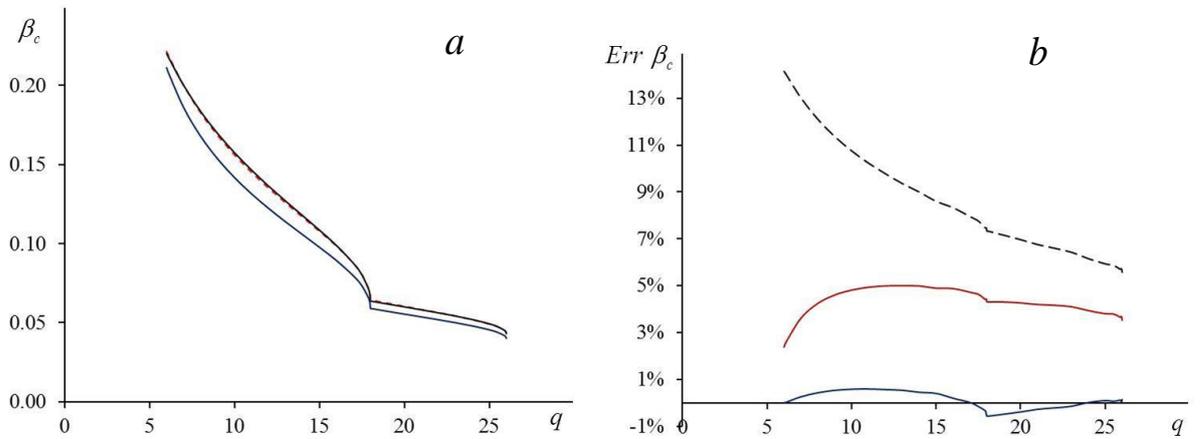

Fig. 3. (a) Dependence of critical temperature $\beta_c = \beta_c(q)$ on effective number of neighbors $q$ for three-dimensional Ising model: lower curve corresponds to Gaussian approximation (see Eq. (18)); upper solid line presents experimental data for $L = 64$; dashed line that merges with upper solid line is calculation with account for third moment (see Eq. (42)). (b) Relative error: upper solid line corresponds to Gaussian approximation (18); lower solid line corresponds to account for third moment; dashed line is a comparison of experiment and mean field theory.

### B. Dependence $C = C(\beta, q)$

Our analysis shows that in the case of a three-dimensional Ising model our account for the next-nearest and the next-next-nearest neighbors does not qualitatively change the behavior of the heat capacity at the critical point. Next, when the number of spins $N$ increases, the peak of the curve $C = C(\beta, q)$ only increases and the character of the singularity remains the same for all $6 \leq q \leq 26$.

The experimental dependences of the heat capacity on the critical temperature differ significantly from the ones defined by the equations (19) and (23). This means that we cannot use Eqs. (21) – (23) to describe the dependences $C = C(\beta, q)$. The equation (25) that defines the finite jump of the heat capacity and works well when $d > 4$, is not applicable in the three-dimensional case.

In Fig. 4, we show the dependences of the maximal energy variance $\sigma^2_{max} = \sigma^2_{max}(q)$ on the effective number of neighbors. From the figure it is obvious that, first, the energy variance at the critical point increases when $q$ increases. Second, we see that with an increase of the lattice size $L$ the difference between the experiment and the theoretical curve grows quickly. Nevertheless, for the lattices of the finite sizes the dependences of $\sigma^2_{max} = \sigma^2_{max}(q)$ on $q$ repeat qualitatively the theoretical curve (compare solid and dashed lines in Fig. 4). Let us note that "the Gaussian" finite value of the critical heat capacity (24) does not reproduce the infinite discontinuity that is the known behavior of this characteristic.

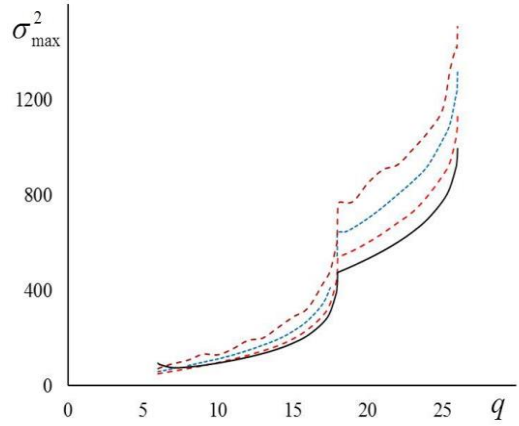

Fig. 4. Dependence $\sigma^2_{max}$ vs $q$: solid line – Eq. (24); dashed lines – computer simulations for $L = 20, 32, 64$ (bottom-to-top).

The situation with the lattices of larger dimensions $d$ is quite the opposite. Let us discuss it here shortly. The larger the dimension of the lattice the better the theoretical expressions (21) – (23) describe the experimental results for the heat capacity in the vicinity of the critical point. When $d \geq 5$, the agreement is good both qualitatively and quantitatively.

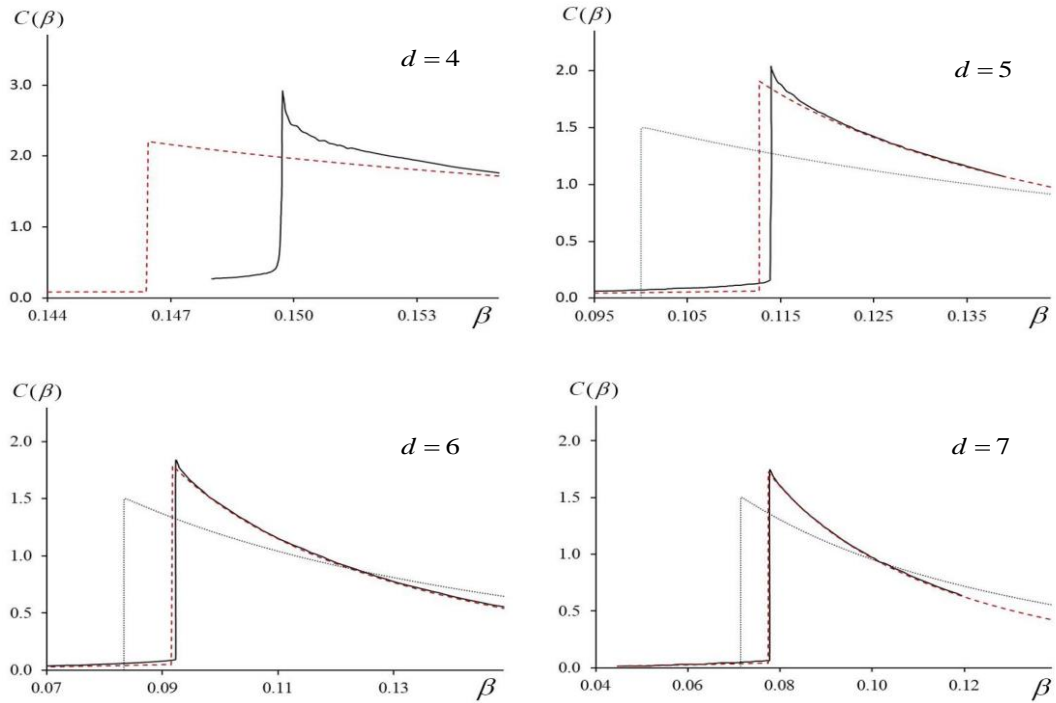

Fig. 5. Heat capacity $C$ vs $\beta$ for $d$ = 4, 5, 6, and 7. Experiment - solid lines; theory (Eqs. (19) and (23)) - dashed lines. Data for $d = 4$ and $L = 80$ are from [21]; curves for $d$ = 5, 6, and 7 are results of experiment [23]. Dotted line in figures for $d$ = 5, 6, and 7 represents

In Fig. 5, the dashed lines are the theoretical curves $C = C(\beta)$ (Eqs. (21) – (23)); the solid lines are the results of computer simulations: for $d = 4$ the value of $L = 80$ (see [21]), for the dimensions $d$ = 5, 6, and 7 we use the data from [23]. We see that our formulas describe the results of simulations pretty well. When $d$ increases, the agreement of the theory and the experiments improves notably; and when $d = 7$ the theoretical and the experimental curves almost merge. In the same figure, we also present the graph for $d = 4$, however up to now there is no real understanding what happens in the four-dimensional case. We do not know for sure if there is a jump of the heat capacity or an infinite singularity. In the same figures, we show the curves obtained in the framework of the mean field theory. We see that our approximation describes the results of the experiments much better.

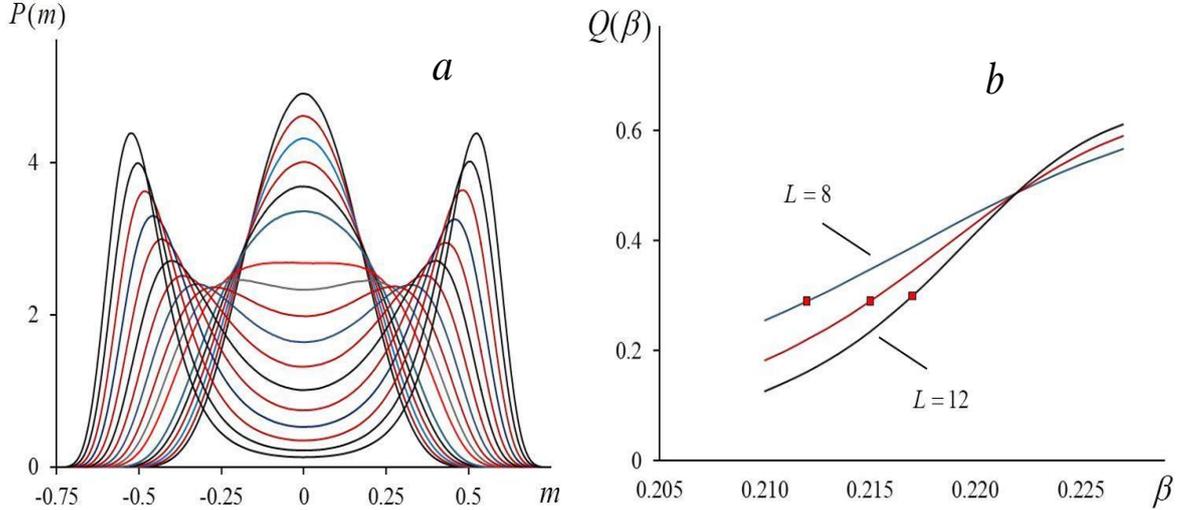

Fig. 6. Magnetization: (a) distribution of magnetization $P(m)$ for three-dimensional Ising model ($q = 6$, $L = 12$) for different inverse temperatures $\beta = 0.210, 0.211, ..., 0.227$; (b) dependence of Binder cumulant on $\beta$ for three-dimensional lattices with $L = 8, 10,$ and $12$. Markers show values of $\beta$ where bimodal distribution replaces unimodal.

### C. Magnetization distribution $P(m)$

With the aid of the Metropolis algorithm we calculated the magnetization distribution near the critical temperature for the three-dimensional Ising model taking into account the nearest neighbors only (L=8, 10, and 12). To increase the accuracy of the Monte Carlo method we performed $3 \cdot 10^7 \cdot L^3$ steps. To fix the inverse temperature where the bimodal distribution replaces the unimodal, we used the step size $\Delta\beta = 10^{-3}$. For $L$=12, we present the curves of the magnetization distribution in Fig. 6a. The obtained distributions allow us to calculate the Binder cumulants $Q(\beta)$ (see [10] and [31]). In Fig. 6b, the curves $Q = Q(\beta)$ intersect at a point $\beta = 0.222$, $Q = 0.488$. This value of $\beta$ defines the critical temperature for the infinite lattice. At the curves in Fig. 6b, the square markers are the points where the bimodal distributions replace the unimodal. At these points, the cumulants are approximately equal to 0.29, and this value is sufficiently close to the value predicted by Eq. (32). When $L$ increases, these points shift to the point of the curves intersection that corresponds to the critical value $\beta_c$ in the asymptotic limit $L \to \infty$.

### D. Logarithmic density of states

To obtain the density of states we used the algorithm of Wang and Landau [11]. We performed the calculations for the cubic lattice of the size $L = 40$ without parallel computing and only accounting for the interaction with the nearest neighbors. As a criterion of the flatness of the histogram of visited states, we adopted the condition that all the values of the histogram had to be larger than 80% of its mean value. When this condition was satisfied, the algorithm reduced the modification factor according to the formula $f_{i+1}^{(mod)} = \sqrt{f_i^{(mod)}}$. The simulations stopped when the modification factor became less than $f_{final}^{(mod)} = \exp(10^{-10})$.

In Fig. 7a, we present the graphs of the experimental logarithmic density of states as well as the ones based on the equations of Sec. II. The density of states calculated using equations (19) – (23) and the Legendre relations was in a good agreement with the experimental data. We see that the maximal obtained error, which is about 0.7 %, corresponds to the critical energy $E = U_c$. The results of the approximate formula (34) are somewhat less accurate; however, in these calculations the deviation from the experimental results is also less than 0.8% (see Fig. 7b).

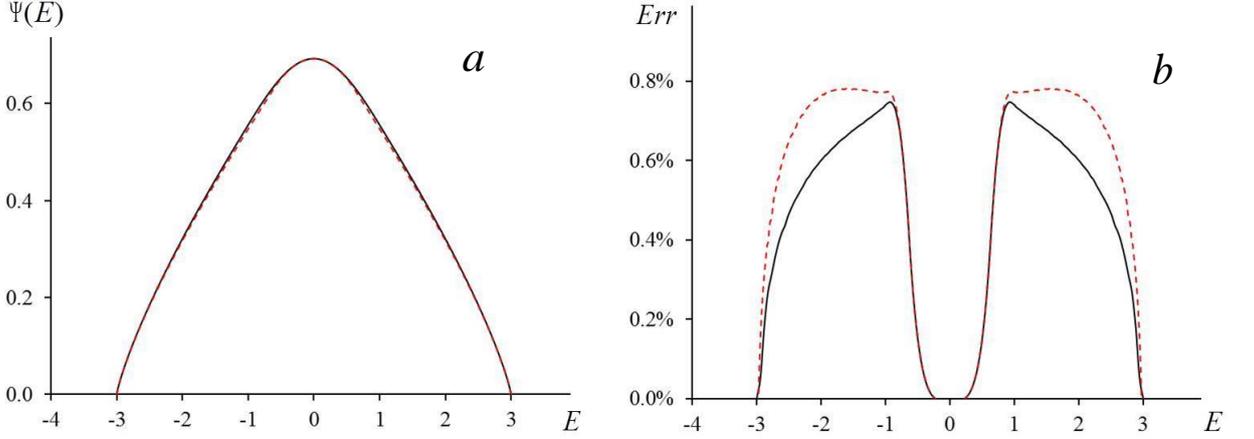

Fig. 7. (a) Logarithmic density of states: solid line - Wang-Landau algorithm for cubic lattice of size $L=40$ ($q = 6$); dashed line - calculation by formulas (19 – 23) and Legendre relations. (b) Solid line is modulus of relative error for calculations when using Legendre relations; dashed line is relative error of Eq. (34).

## VII. DISCUSSION

We used the *m*-vicinity method to investigate the Ising Model on *d*-dimensional hypercube lattices for $3 \leq d \leq 7$. (We recall that this method is not applicable for the lattices of lower dimensions.) The *m*-vicinity $\Omega_m$ consists of all the configurations of the same magnetization *m*. We based our method on the series expansion of the logarithm density of states in the *m*-vicinities (Sec. II). In the main part of the paper, we discuss the interaction with the nearest neighbors. Only in Sec. VI, for a cubic lattice we analyzed both the short-range and long-range interactions.

When we account only for an isotropic interaction of the nearest neighbors, the small parameter is $|\varepsilon_{\max}| = \sqrt{d}\left(1 - \sqrt{1 - 2/d}\right)/2$ (see Eq. (10)). The value of this parameter decreases from 0.366 when $d = 3$ to 0.159 when $d = 7$ (see Fig. 1). Not surprisingly, the agreement between the theoretical results and computer simulations becomes better when the dimension of the lattice increases.

Summing up, we would like to list the main features of the *m*-vicinity method.

*1*) The first order of the perturbation theory is equivalent to the Gaussian approximation of the true density of states in the *m*-vicinities supposing that the first two moments of the density of states and its Gaussian approximation coincide.

In this case, we obtained the simple analytical expression (18) for the critical value of the inverse temperature, which described quite accurately the results of computer simulations for different lattices (see Fig. 2). When $d = 3$, the relative error between the theoretical estimate $\beta_c$ and the experiment is 2.39%. When *d* increases, the relative error decreases to 0.18% for $d = 7$ (see Table 1).

In the framework of the same approximation, we obtained analytical expressions (19) – (23) that define implicitly the dependence of the free energy and its derivatives on the temperature. Based on these results we calculated the dependence of the heat capacity $C(\beta)$ on the inverse temperature $\beta$. As it follows from Fig. 5, for $d \geq 5$ the obtained graphs match the experimental curves [23].

The expressions for the critical parameters that follow from Eqs. (25) – (28) coincide with the known mean field results. For $d \geq 5$, the equation (25) predicts a jump of the heat capacity whose value is only 0.6% less than the value obtained by computer simulations [23].

*2*) The second order of the perturbation theory requires going beyond the Gaussian approximation and account for the third moment in the expansion (36). We suppose that the first three moments of the approximate and the true distributions coincide. However, this does not automatically improve the agreement of the theoretical estimates of the critical temperature and its experimental values. The reason is an irregular convergence of the Gram-Charlier and the Edgeworth series expansions [24]. The agreement improves significantly when we introduce the adjustable parameter that is the same for all the dimensions $d$ (see Sec. V). After that, the second order perturbation formula (41) describes the experiment within a fraction of the percent (see Table 1). From Fig. 3 it follows that the same is true when we also account for the interactions with the next-nearest and the next-next-nearest neighbors.

*3)* The results of our analysis show that when the dimension of the Ising model $d \geq 5$, the *m*-vicinity method describes the properties of the system pretty good, both qualitatively and quantitatively. When $d = 4$, the type of the singularity of heat capacity is still remains unclear [20]-[22]. Consequently, the question about the applicability of the *m*-vicinity method in this case remains open. In the case $d = 3$, the approach discussed here is incorrect because it predicts a finite jump of the heat capacity at the critical point. Nevertheless, this method allows one to calculate the critical temperature quite accurately (see Table 1) as well as to describe its dependence on the number of the neighbors (see Fig. 3 and Eq. (57)). We conclude that when $d = 3$ our theory provides good results for the dependences of the free energy and the logarithmic density of states but not for their derivatives. Indeed, when for $d = 3$ we use the equations (34) and (35) to calculate the logarithmic density of states, the result is in a good agreement with the experimental data. From Fig. 7 we see a notable deviation from the experiment ($\sim 0.7\%$) only in a narrow vicinity of the point $E = U_c$.

*4)* Finally, we would like to note that a good agreement of the theoretical results for the density of states $\Psi = \Psi(E)$ with the data of computer simulations allows us to use the obtained expressions as an initial approximation for the Wang-Landau algorithm. We hope that our results will allow one to speed the algorithm up and to increase its accuracy.

## ACKNOWLEDGMENTS

The work was done in the framework of the State program of SRISA RAS No. 0065-2019-0003.

## APPENDIX. CALCULATION OF THE FIRST MOMENTS OF ENERGY DISTRIBUTION

Previously, we described in detail how to calculate the moments of the distribution of the energy states belonging to the vicinity $\Omega_m$ [26] – [28]. In these papers, we obtained the exact expressions for the first two moments of the energy distribution that are the mean $E_m$ and the variance $\sigma_m^2$. For a finite system and a general connection matrix $\mathbf{J} = \left( J_{ij} \right)_{i,j=1}^{N}$, we presented these expressions in finitely combinatorial forms. In the case of the Ising connection matrix for a *d*-dimensional hypercube, we in addition found $E_m$ and $\sigma_m^2$ in an asymptotic limit when the number of spins $N$ tends to infinity. In what follows we show how to obtain the combinatorial and the asymptotic formulas for the third moment $\mu_3$.

### A. Notations

In what follows, for the sake of simplicity when averaging over the *m*-vicinities $\Omega_m$, we sometimes will omit the subscript *m*. For example, if a state $\mathbf{s} = (s_1, s_2, ..., s_N)$ belongs to an *m*-vicinity $\Omega_m$ and $E(\mathbf{s})$ is the energy of the state $\mathbf{s}$, we will write the first three moments as

$$\langle E \rangle = N_m^{-1} \sum_{\mathbf{s} \in \Omega_m} E(\mathbf{s}), \quad \langle E^2 \rangle = N_m^{-1} \sum_{\mathbf{s} \in \Omega_m} E^2(\mathbf{s}), \quad \text{and} \quad \langle E^3 \rangle = N_m^{-1} \sum_{\mathbf{s} \in \Omega_m} E^3(\mathbf{s}), \quad \text{where} \quad N_m \equiv \binom{N}{n}.$$

We are interested in calculating the second and third central moments – the variance $\sigma^2$ and the semi-invariant $\mu_3$:

$$\mu_3 = \langle E^3 \rangle - 3\sigma^2 \langle E \rangle - \langle E \rangle^3, \quad \text{and} \quad \sigma^2 = \langle E^2 \rangle - \langle E \rangle^2.$$

Usually, the energy $E(\mathbf{s}) = a \cdot \sum_{i,j=1}^{N} J_{ij} s_i s_j$ of the state $\mathbf{s}$ includes a normalization coefficient $a$. In what follows, we for simplicity omit this coefficient and write the energy as

$$E(\mathbf{s}) = \mathbf{s}\mathbf{J}\mathbf{s}^+ \equiv \sum_{i,j=1}^{N} J_{ij} s_i s_j . \tag{A1}$$

After obtaining the final expressions for the moments, we have to multiply $\langle E \rangle$, $\sigma^2$, and $\mu_3$ by $a$, $a^2$, and $a^3$, respectively.

### B. Connection matrix of general form

Suppose that the connection matrix has a general form and the configuration $\mathbf{s}_0 = (s_{01}, s_{02}, ..., s_{0N})$ is the ground state whose energy is $E_0 = \mathbf{s}_0 \mathbf{J} \mathbf{s}_0^+$.

To calculate the energy moments $\langle E^r \rangle$, $r = 1, 2, ...$, it is necessary to present $E^r$ as a sum of terms which contain only the products of spin variables with non-repeating indices. For example, we can write the expressions for $E$ and $E^2$ as

$$E = \sum_{i=1}^{N} \sum_{j=1}^{N} J_{ij} s_i s_j \chi_{ij}$$

$$E^2 = 2\operatorname{Tr} \mathbf{J}^2 + 4\sum_{i,j=1}^{N} (\mathbf{J}^2)_{ij} s_i s_j \chi_{ij} + \sum_{i,j,k,r=1}^{N} J_{ij} J_{kr} s_i s_j s_k s_r \chi_{ijkr} ,$$

where $\chi_{ij}$, and $\chi_{ijkr}$ are tensors whose components are nonzero and equal to one only if they do not contain the repeating indices.

In the $m$-vicinity we can present the variables $s_i$ as $s_i = s_{0i} p_i$, where $p_i$ takes the value $p_i = -1$ with the probability of $n/N$ and it takes the value $p_i = 1$ with the probability $(N-n)/N$. (We remind that $n = N(1-m)/2$). This allows us to easily perform the averaging over the $m$-vicinities. For example:

$$\langle E \rangle = N_m^{-1} \sum_{\Omega_m} \sum_{i=1}^{N} \sum_{j=1}^{N} J_{ij} s_{0i} s_{0j} p_i p_j \chi_{ij} = \sum_{i=1}^{N} \sum_{j=1}^{N} J_{ij} s_{0i} s_{0j} \cdot N_m^{-1} \sum_{\Omega_m} p_i p_j \chi_{ij} = E_0 K_2 \tag{A2}$$

When writing Eq. (A2) we used easily verifiable formulas $N_m^{-1} \sum_{\Omega_m} p_i p_j \chi_{ij} = K_2 \chi_{ij}$, $N_m^{-1} \sum_{\Omega_m} p_i p_j p_k p_r \chi_{ijkr} = K_4 \chi_{ijkr}$, and $N_m^{-1} \sum_{\Omega_m} p_i p_j p_k p_r p_m p_n \chi_{ijkrmn} = K_6 \chi_{ijkrmn}$:

$$\begin{aligned} K_2 &= \frac{(N-2n)^2 - N}{N(N-1)}, \\ K_4 &= \frac{(N-2n)^4 - 2(N-2n)^2(3N-4) + 3N(N-2)}{N(N-1)(N-3)(N-3)}, \\ K_6 &= 1 - \frac{4(N-n) \cdot n \cdot B}{N(N-1)(N-2)(N-3)(N-4)(N-5)}. \end{aligned} \tag{A3}$$

In Eq. (A3), the constant $B$ in the numerator of the last expression is equal to

$$\begin{aligned} B = &\, 3(N-2n)^4 + 6(N-2n)^3 \cdot (2n-5) + (N-2n)^2 \cdot (28n^2 - 120n + 125) + \\ &+ (N-2n)(32n^3 - 180n^2 + 340n - 210) + 4(n-1)(n-2)(4n^2 - 18n + 23). \end{aligned}$$

The same as when calculating $<E>$ in Eq. (A2), we can obtain the second and the third energy moments and show that the averaging over $\Omega_m$ leads to the following equalities

$$E_m = K_2 E_0,$$
$$\sigma_m^2 = 2(1 - 2K_2 + K_4) \cdot \text{Tr} \mathbf{J}^2 + 4(K_2 - K_4) \cdot \left( \mathbf{s}_0 \mathbf{J}^2 \mathbf{s}_0^+ \right) - (K_2^2 - K_4) E_0^2, \tag{A4}$$
$$\mu_3(m) = 8\left(1 - 3K_2 + 3K_4 - K_6\right) \cdot \text{Tr} \mathbf{J}^3 - 16(K_4 - K_6) \cdot \sum_{i=1}^{N} s_{0i} h_{0i}^3 -$$
$$- 12 E_0 \left( \mathbf{s}_0 \mathbf{J}^2 \mathbf{s}_0^+ \right)(K_2^2 - K_4 - K_2 K_4 + K_6) +$$
$$+ 6 E_0 \text{Tr} \mathbf{J}^2 \left[ K_6 - K_2 K_4 + 2(K_2^2 - K_4) \right] + E_0^3 \left( K_6 - 3K_2 K_4 + 2K_2^3 \right) +$$
$$+ 8\left( K_2 - 2K_4 + K_6 \right) \left[ 2 \left( \mathbf{s}_0 \mathbf{J}^{(3)} \mathbf{s}_0^+ \right) - 6 \sum_{i=1}^{N} s_{0i} h_{0i} \cdot \left( \mathbf{J}^2 \right)_{ii} + 3 \left( \mathbf{s}_0 \mathbf{J}^3 \mathbf{s}_0^+ \right) \right],$$

where $h_{0i} = \sum_{j=1}^{N} J_{ij} s_{0j}$ is a local field acting on the *i*-th spin belonging to state $\mathbf{s}_0$. The matrix $\mathbf{J}^{(3)}$ in the expression for the third moment is $\mathbf{J}^{(3)} = \left( J_{ij}^3 \right)$.

The expressions (A3) and (A4) provide us with the most general formulas for the first, second, and third moments of the distribution of energies of the states from the *m*-vicinity $\Omega_m$ in terms of the elements of the connection matrix $\mathbf{J}$. (If it is necessary to use the normalized expression for the energy $E(\mathbf{s})$, see the note after Eq. (A1).)

### C. Ising model on hypercube

In this subsection, we analyze the Ising model with the short-range interaction, when only the nearest spins interact and other elements of the connection matrix are equal to zero. Let the nonzero matrix elements be equal to a constant $J$. We factor out this constant in all our calculations. Then there is $q = 2d$ ones in each row of the matrix $\mathbf{J}$ and all the other matrix elements are equal to zero and the ground state is a configuration $\mathbf{s}_0 = (1,1,1,...,1) \in \mathbb{R}^N$.

It is easy to see that the following equations

$$E_0 = qN, \quad \text{Tr} \mathbf{J}^2 = qN, \quad \mathbf{s}_0 \mathbf{J}^2 \mathbf{s}_0^+ = q^2 N, \quad \text{Tr} \mathbf{J}^3 = 0,$$
$$\sum_{i=1}^{N} s_{0i} h_{0i} \cdot \left( \mathbf{J}^2 \right)_{ii} = q^2 N, \quad \mathbf{s}_0 \mathbf{J}^3 \mathbf{s}_0^+ = q^3 N, \quad \sum_{i=1}^{N} s_{0i} h_{0i}^3 = q^3 N.$$

hold. Then the expressions (A4) take the form

$$E_m = qN \cdot K_2$$
$$\sigma_m^2 = qN \cdot \left\{ qN \cdot (K_4 - K_2^2) + 2\left[ 1 - 2K_2 + K_4 + 4q(K_2 - K_4) \right] \right\} \tag{A5}$$
$$\mu_3(m) = (qN)^3 \left( K_6 - 3K_2 K_4 + 2K_2^3 \right) +$$
$$+ 6(qN)^2 \left[ K_2^2 - K_4 + (1 - 2q)\left( K_2^2 - K_4 + K_6 - K_2 K_4 \right) \right] +$$
$$+ 8(qN) \left[ \left( K_2 - K_4 \right)\left( 2 - 6q + 3q^2 \right) + \left( K_6 - K_4 \right)\left( 2 - 6q + 5q^2 \right) \right].$$

To obtain asymptotic expressions for the moments ($N \to \infty$), at first it is necessary to obtain the asymptotic expressions for the coefficients $K_2$, $K_4$, and $K_6$ expanding them with respect to the small parameter $\delta = 1/N$:

$$K_2 \approx m^2 - \delta \cdot (1 - m^2) - \delta^2 (1 - m^2),$$
$$K_4 \approx m^4 - \delta \cdot 6m^2 (1 - m^2) + \delta^2 (1 - m^2)(3 - 25m^2),$$
$$K_6 \approx m^6 - \delta \cdot 15 m^4 (1 - m^2) + \delta^2 5 m^2 (1 - m^2)(9 - 28 m^2).$$

Here $m = (N - 2n)/N \in [-1, +1]$ is the magnetization of the state $\mathbf{s} \in \Omega_m$.

Next, substituting the obtained expressions into Eqs. (A5) and leaving the leading terms we obtain the asymptotic expressions for the first three moments

$$E_m = N \cdot qm^2,$$
$$\sigma_m^2 = N \cdot 2q\left(1 - m^2\right)^2, \tag{A6}$$
$$\mu_3(m) = N \cdot 16 q m^2 \left(1 - m^2\right)^2.$$

The final answer we obtain after multiplying $E_m$, $\sigma_m^2$, and $\mu_3(m)$ by $a$, $a^2$, and $a^3$, respectively; here $a = -J/2N$.

**D. Asymptotic form of coefficient $\kappa_3$**

Starting from Eq. (8) we used the series expansion of the unknown function $\varphi(m, E)$ in the variable $\varepsilon = (E - E_m)/\sigma_m$:

$$\varphi = \frac{1}{2}\varepsilon^2 + \frac{1}{3!}\kappa_3\varepsilon^3 + \frac{1}{4!}\kappa_4\varepsilon^4 + \dots . \tag{A7}$$

Since we do not know the function $\varphi(m, E)$ itself, we also do not know the series expansion coefficients $\kappa_r$, where $r = 3, 4, \dots$. However, we do know the first three moments of the energy distribution (A5) and this allows us to define the coefficient $\kappa_3$ of the expansion (A7). Indeed, Eqs. (A5) and (A6) define the general form of the third moment:

$$\mu_3(m) = \int (E - E_m)^3 \exp[-N \cdot \varphi(m, E)] dE . \tag{A8}$$

We restrict ourselves with the first two terms in the right-hand side of Eq. (A7) and substitute the expressions (A5) – (A7) into the integral in the right-hand side of Eq. (A8). Then

$$\mu_3(m) = \sigma_m^4 \int \exp\left[-N \cdot \left(\frac{1}{2}\varepsilon^2 + \frac{1}{3!}\kappa_3\varepsilon^3\right)\right] \varepsilon^3 d\varepsilon . \tag{A9}$$

It is easy to see that the main contribution to the integral (A9) comes from the region $\varepsilon \sim \pm N^{-1/2}$, where the cubic term of the exponent of the integrand is negligible small $(N\varepsilon^3 \sim N^{-1/2})$. Then expanding the exponent we have

$$\mu_3(m) = \sigma_m^4 \int \exp\left(-\frac{1}{2}N\varepsilon^2\right) \cdot \left(1 - \frac{1}{3!}N\kappa_3\varepsilon^3\right) \varepsilon^3 d\varepsilon . \tag{A10}$$

After simple calculations, equating the expression (A10) and the third of the expressions (A6), we obtain with the accuracy up to the terms of the order of $\sim O(N^{-1})$

$$\kappa_3 = -\frac{\mu_3(m)}{\sigma_m^3} = \frac{2qm^2}{\sigma_0^3(1-m^2)} . \tag{A11}$$

In the same way, it is also possible to calculate the coefficients of the higher order. However, even the expression for $\kappa_4$ is so cumbersome that we do not present it here.

**References**

[1]    B. Müller, J. Reinhardt, M. T. Strickland, *Neural Networks - An Introduction* (Berlin, Springer Verlag, 1995).

[2]    V. Dotsenko, *An introduction to the theory of spin glasses and neural networks* (Singapore, World Scientific, 1994).

[3]    Ia. Karandashev, B. Kryzhanovsky and L. Litinskii, Phys. Rev. E **85**, 041925 (2012).

[4]    Y. Fu, P.W. Anderson, J. Phys. A **19**, 1605 (1986).

[5]    Theoretical Computer Science **265**, 3 (2001).

[6]    A.K. Hartmann, M. Weight, *Phase transitions in combinatorial optimization problems* (Weinheim: WILEY-VCH Verlag GmbH & Co, 2006)

[7]    V.M. Yakovenko, J.B. Rosser (Jr.), Rev. Modern Phys. **81**, 1703 (2009).

[8]    S. Galam, *Sociophysics: A Physicist's Modeling of Psycho-Political Phenomena* (New York, Springer, 2012).

[9]     R. Häggkvist, et al., *Advances in Phys.* **56** (5), 653 (2007).

[10]   D.P. Landau, K. Binder, *A guide to Monte Carlo simulations in statistical physics* (Cambridge, Cambridge Univ. Press, 2000).

[11]   F. Wang, D. P. Landau, Phys. Rev. Lett. **86**, 2050 (2001).

[12]   A.Z. Patashinskii, V.L. Pokrovskii, *Fluctuation theory of phase transitions* (Oxford, Pergamon Pr., 1979).

[13] R. J. Baxter, *Exactly solved models in statistical mechanics* (London, Academic Press, 1982).

[14] A. Dmitriev, V. Katrakhov, Yu. Kharchenko, *Root transfer matrices in Ising Models* (Moscow, Nauka, 2004, in Russian)

[15] J.M. Dixon, J.A. Tuszynski, E.J. Carpenter, Phys. A **349**, 487 (2005).

[16] A. Ferrenberg, J.Xu, D.P. Landau, Phys. Rev. E **97**, 043301 (2018).

[17] D. Roon, A. Brandt, R.H. Swendsen, Phys. Rev. E **95**, 053305 (2017).

[18] P. Butera, M. Comi, Phys. Rev. B **65**, 144431 (2002).

[19] A. Murtazaev et al., JETP **120**, 110 (2015).

[20] H. J. Blote and R. H. Swendsen, Phys. Rev. B **22**, 4481 (1980).

[21] P.H. Lundow, K. Markstrom, Phys. Rev. E **80**, 031104 (2009).

[22] S. Akiyama et al. Preprint arXiv: 1911.12954v1.

[23] P.H. Lundow, K. Markstrom, Nucl. Phys. B **895**, 305 (2015).

[24] H. Cramer, *Mathematical methods of statistic* (Princeton, Princeton Univ. Press, 1999).

[25] M.G. Kendall, A. Stuart, *The advanced theory of statistic: Distribution Theory*. (London, Charles Griffin & Comp. Lim., 1958).

[26] B.V. Kryzhanovsky, L.B. Litinskii, Doklady Mathematics **90**, 784 (2014).

[27] B. Kryzhanovsky, L. Litinskii, Physica A **468**, 493 (2017).

[28] B. Kryzhanovsky, L. Litinskii, Opt. Mem. and Neur. Nets. **24**, 165 (2015).

[29] W.L. Bragg, E.J. Williams. Proc. Roy. Soc. **A145**, 699 (1934).

[30] B.V. Kryzhanovsky, Doklady Physics **64**, 280 (2019).

[31] K. Binder, Z. Phys. B **43**, 119 (1981).

[32] Rosana A. dos Anjos et al., Phys Rev E **76**, 022103 (2007).

[33] A.K. Murtazaev et al., Phys. of Sol. St. **59**, 110 (2017).